\documentclass[a4paper,11pt]{article}
\pdfoutput=1
\usepackage{jheppub}
\usepackage{slashed}
\usepackage[T1]{fontenc}
\usepackage{adjustbox}
\usepackage{multirow}
\usepackage{hhline}
\usepackage{tabularx,booktabs,caption}
\usepackage{float}
\usepackage{placeins}
\usepackage{caption}
\usepackage{verbatim}
\usepackage{soul}
\usepackage[utf8]{inputenc}
\usepackage{amsmath,etoolbox}
\AtBeginEnvironment{pmatrix}{\setlength{\arraycolsep}{10pt}}

\usepackage{tikz}
\usepackage{cancel}
\usepackage{graphicx}

\usepackage{xcolor}
\usepackage{hyperref}
\usepackage{dirtytalk}

\newcommand{\beq}{\begin{equation}}
\newcommand{\eeq}{\end{equation}}
\newcommand{\beqn}{\begin{eqnarray}}
\newcommand{\eeqn}{\end{eqnarray}}
\newcommand{\bea}{\begin{eqnarray}}
\newcommand{\eea}{\end{eqnarray}}

\usepackage{amsmath}
\usepackage{amssymb}
\usepackage{comment}
\usepackage{slashed}
\usepackage[normalem]{ulem}
\usepackage{cancel}
\usepackage{dirtytalk}
\usepackage{empheq}
\usepackage{multirow}
\usepackage{graphics}

\newcommand*\widefbox[1]{\fbox{\hspace{2em}#1\hspace{2em}}}

\title{\boldmath $\sin^2\theta_W$ and neutrino electromagnetic interactions in CE$\overline{\nu}_e$NS with different quenching factors}

\author{Amir N.\ Khan,\note{Corresponding author.}}
\affiliation{ Max-Planck-Institut f\"ur Kernphysik, Postfach
103980, D-69029 Heidelberg, Germany} 

\emailAdd{amir.khan@mpi-hd.mpg.de, ntrnphysics@gmail.com}

\abstract{Recently, evidence for the observation of about 2 keV and below nuclear recoils from the coherent scattering of reactor anti-neutrinos off the germanium nuclei has been reported. We analyze the observed data to estimate the value of the weak mixing angle and constrain the neutrino millicharge, magnetic moment, charge radius and anapole moment contributing to the coherent scattering process. Currently, there is no definite model available for the quenching factor at such low energies. To this end, we consider various models of the quenching factor and show how it affects the interpretation of the obtained results. We find that the bounds obtained are stronger in some cases while comparable or weaker in other cases which show a strong dependence on the choice and accuracy of a particular quenching factor model. The results are the first at such low-energy nuclear recoils. We present an exhaustive list of analytical functions for the different quenching factors corresponding to the existing models and to the data from various experiments. Such functions will be useful for any new physics study using the nuclear recoils due to the reactor neutrinos.}

\begin{document} 
\captionsetup[figure]{labelfont={bf},labelformat={default},labelsep=period,name={FIG.}}
\maketitle

\section{Introduction}\label{sec:intro}
\section{Introduction}

The first evidence for the observation of coherent elastic anti-neutrino nucleus scattering (CE$\overline{\nu}_e$NS) with Dresden-II boiling water reactor neutrinos and p-type point-contact germanium, also known as NCC-1701 detector, has been suggested in refs. \cite{Colaresi:2022obx,Collar:2022ibv}, following from the earlier work \cite{Colaresi:2021kus}. The result also suffers from large experimental uncertainties in the region of interest (sub-keV). On top of that, there is no unique theoretical prediction of the standard event spectrum possible because of our limited understanding of the quenching factor (QF) at such low energies. This leads to a large disparity in new physics predictions. The maximum nuclear recoil produced by reactor neutrinos is up to $2 \ \rm keV$ and the corresponding ionization energy is below one keV, while the detection threshold is 0.2 keV ionization energy. Other competitive experiments looking for nuclear recoils with reactor neutrinos such as MINER\cite{Agnolet:2016zir}, RED-100 \cite{Akimov:2017hee}, $\nu$-cleus \cite{Strauss:2017cuu}, CONUS \cite{Hakenmuller:2019ecb, CONUS:2020skt, CONUS:2021dwh,Bonhomme:2022lcz,CONUS:2022qbb}, CONNIE \cite{CONNIE:2019swq,CONNIE:2019xid,CONNIE:2021ngo}, TEXONO \cite{TEXONO:2020vnv}, vIOLETA \cite{Fernandez-Moroni:2020yyl} and SCB \cite{SBC:2021yal} will provide further concrete information about this observation.

The coherent elastic neutrino-nucleus scattering process was predicted long ago \cite{Freedman:1973yd, Freedman:1977xn,Tubbs:1975jx,Drukier:1984vhf}, however, it was first observed only a few years ago by the COHERENT experiment with nuclear recoils of 10 keV and above and with neutrino energies of a few tens of MeV produced from spallation neutron source \cite{Akimov:2017ade, Akimov:2018vzs, Akimov:2021dab}. Several standard model (SM) and nuclear physics parameters were studied \cite{Barranco:2005yy, Cadeddu:2017etk,Tomalak:2020zfh} and new physics interactions and models were explored using the coherent scattering process \cite{Anderson:2012pn,deNiverville:2015mwa,Lindner:2016wff,Coloma:2017ncl,Bauer:2018onh, Billard:2018jnl,Denton:2018xmq, Dutta:2019eml, Khan:2019cvi, Khan:2021wzy, Khan:2022not, Coloma:2022avw, Liao:2022hno, AristizabalSierra:2022axl,Corona:2022wlb,Abdullah:2022zue,delaVega:2021wpx,Denton:2020hop,Suliga:2020jfa,Du:2021idh,Cerdeno:2021cdz,Alikhanov:2021dhb,Fernandez-Moroni:2021nap}. With reactor neutrinos, no signal of the CE$\overline{{\nu}_e}$NS process was observed yet until recently when its observation evidence was reported with Dresden-II reactor neutrinos and NCC-1701 detector \cite{Colaresi:2022obx}. 

Here, we analyze the data of the observed measurement with the 3 kg detector mass, 2.96 Giga-Watt reactor and data-taking period of 96.4 days \cite{Colaresi:2022obx}. We mainly focus on estimating the weak mixing angle ($\sin^2{\theta_W}$) and derive bounds on the electromagnetic interactions; neutrino millicharges, magnetic moment, charge radius and anapole moment. At such low energies, the detection of nuclear recoils suffers from the lack of accurate knowledge about the quenching factor and from large uncertainties about it. Therefore, we use different models for the quenching factor in our analysis and compare our results from them. Two of them are theoretical models; one is the famous Linhard model \cite{lindhard1963integral} while the other is an ansatz introduced by Sarkis et al \cite{Sarkis:2020soy} which is also based on a modification to the Linhard model and an improvement of recent work \cite{Sorensen:2014sla}. In the third case, we will use the experimental data of the QF obtained from the NCC-1701 detector calibration, called Iron-filtered (Fef) data. In all three cases, we evaluate the quenching factor as a function of the observed ionization energy.

We will show how the three quenching factors give different results for $\sin^2{\theta_W}$ and for electromagnetic properties of neutrinos. At such low energies, neutrino millicharge is more sensitive than the other electromagnetic interactions due to their interference with the standard weak interactions and their dependence on the inverse power of the recoil energy and on the target mass \cite{Khan:2020vaf,Khan:2020csx,XENON:2020rca, Khan:2022not}. Further, we will derive constraints on all electromagnetic properties of neutrinos with the observed data.

The paper is organized as follows. In the next section, we discuss different quenching factors that we will use for our analysis. In Sec. \ref{sec:formalism}, we discuss the differential cross-section of the CE$\nu$NS in the SM and all necessary notations. In Sec. \ref{sec:analysis}, we discuss reactor neutrino fluxes and calculate the event energy spectrum. In Sec. \ref{sec:EMprorties}, we introduce the electromagnetic properties of neutrinos and discuss our results in view of the three QF used here. Finally, we summarize and conclude in Sec. \ref{sec:concl}. 

\section{Models and data for the quenching factors}\label{sec:QF}
For ionization detectors, the visible nuclear recoil energy, called the ionization energy ($ E_I$) is always less than the actual nuclear recoil energy ($E_{nr}$) due to the energy loss in exciting the binding atoms. This effect is even stronger at lower energy nuclear recoils, particularly for the nuclear recoils due to reactor neutrinos. The ratio between the two energies is often called the quenching factor, denoted here by $Q$. This could be conveniently taken as a function of $E_{nr}$ or $E_{I}$. We define it as a function of $E_{I}$ in the following,
\begin{equation}
Q(E_I) = \frac{E_I}{E_{nr}}\,.
\label{eq:QF}
\end{equation}%

The commonly used theoretical model for the quenching factor is the so-called Linhard model ~\cite{lindhard1963integral}. Although this model is successful at nuclear recoils above a few keV, it fails to describe the low energy phenomena. For example, this model predicts that for one keV true nuclear recoil energy for a Germanium nucleus is reduced by about 80\%. This reduction is even more for the nuclear recoils below one keV. This limitation is caused by several approximations made in the formulation of the model of the atomic binding energy at low energies. To account for properly treating the atomic binding energy at lower energy recoils some attempts were made in refs. \cite{Sorensen:2014sla} and \cite{Sarkis:2020soy}. The authors of refs. \cite{Sorensen:2014sla, Sarkis:2020soy} have revisited the original Linhard model and modified it for the low energy recoils by relaxing the binding energy approximations which were made in the formulation of the Linhard model. Ref. \cite{Sorensen:2014sla} includes new kinematic effects and shows how they affect the low energy cut-off for the atomic binding energy. This approach has recently been improved further by Sarkis \textit{$et \ al$} \cite {Sarkis:2020soy} by considering a semi-hard sphere interaction model and by solving the original Linhard integral equation and bringing the cut-off on the binding energy up to $\sim$ 200 eV. Another choice for the quenching factor is to use the experimental data obtained directly for the calibration of the particular detector with photon sources from the neutron scattering measurements.

For our analysis here we will consider three different quenching factors: $i)$ Linhard Model $ii)$ Sarkis $et \ al$ model $iii)$ Iron-filtered (Fef) data obtained from the calibration of the Germanium-based NCC-1701 detector. We have calculated the quenching factor as a function of the ionization energy for each case, as discussed in the following.

\begin{figure}[t]
\begin{center}
\includegraphics[width=5in]{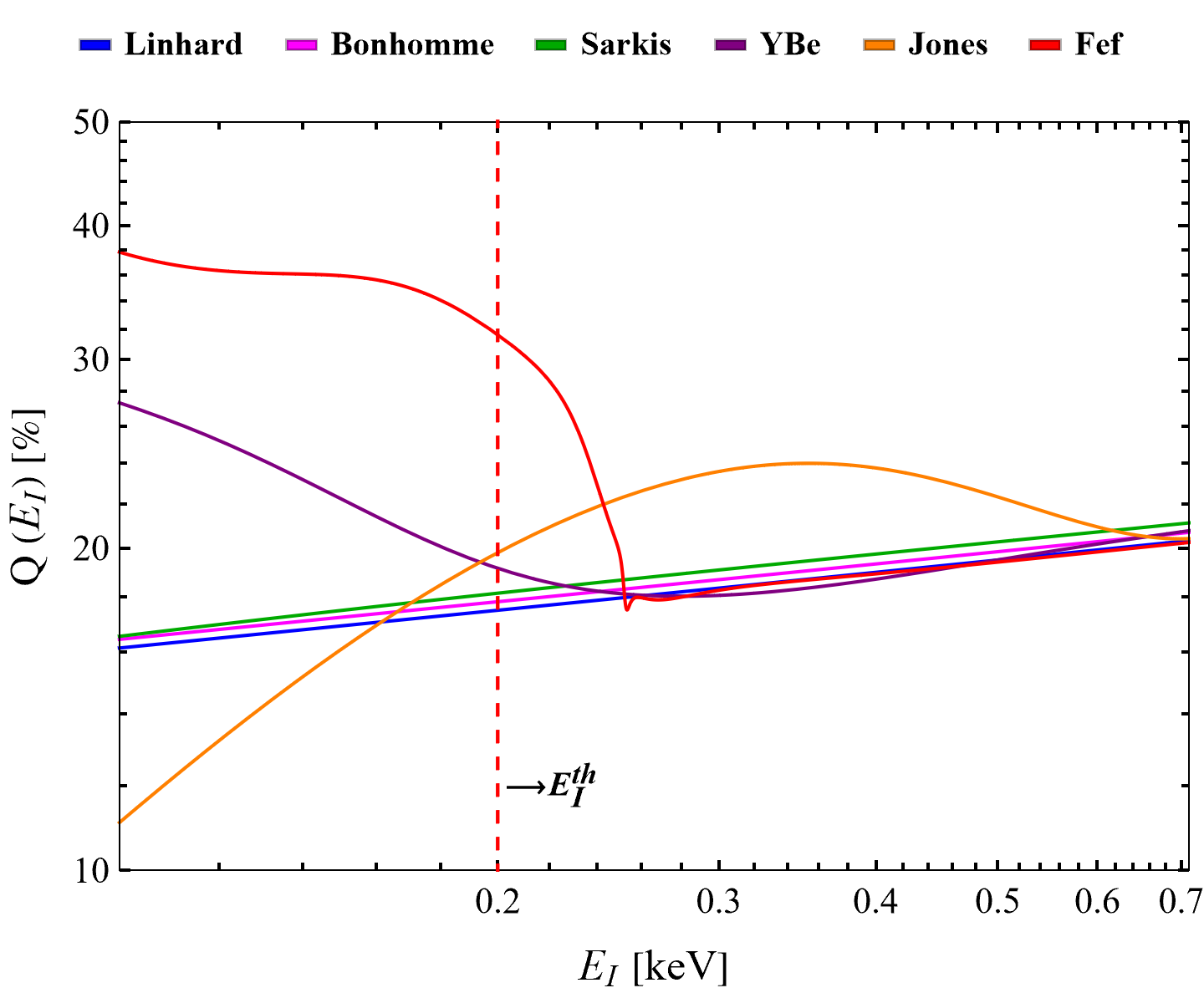}
\end{center}
\caption{The quenching factor models and iron-filtered (Fef) data as a function of the true ionization energy. The dashed vertical line corresponds to the Germanium based NCC-1701 detector threshold of 0.2 keV.}
\label{fig:QF_models}
\end{figure}

\textbf{Linhard Model:} The Linhard model ~\cite{lindhard1963integral} depends on the element-specific parameter ‘$k$’ which is a function of atomic mass and atomic number of the relevant atom $(k = 0.133 \times Z^{2/3} \times A^{-1/2})$. For Germanium $\rm (^{72}Ge)$ with $k = 0.158$, we derive the following quenching factor as a function of the ionization energy,  

\begin{equation}
Q(E_I) = E_I \sum_{i=0}^{6}\left[\frac{a_i E_{I}^i}{b_i E_{I}^i}\right] \ \ \ \ (\rm Linhard \ Model)\,,
\end{equation}
where $a_i$ and $b_i$ are the evaluated fitting parameters. The numerical values of the fit parameters are given in the Ist and 2nd columns of Table \ref{tab:-QF-models}. The resultant quenching factor as a function of $E_I$ is shown in Fig. \ref{fig:QF_models} in blue.

\textbf{Bonhomme data:} The most recent direct measurement of the quenching factor has been reported in \cite{Bonhomme:2022lcz}. Four data sets were obtained at different energies in this measurement and the data were fitted with the Linhard model. The best-fit value of the parameter ‘$k= 0.133$’ was obtained \cite{Bonhomme:2022lcz}. We use the Linhard model with this value and derive the quenching factor in terms of the ionization energy. We obtain the following fit function for the quenching factor as a function of the ionization energy
\begin{equation}
Q(E_I) = E_I \sum_{i=0}^{6}\left[\frac{a_i E_{I}^i}{b_i E_{I}^i}\right] \ \ \ \ (\rm Bonhomme  \ \textit{et al} \ \ data)\,,
\end{equation}
where $a_i$ and $b_i$ are the evaluated fitting parameters. The numerical values of the fit parameters are given in the 3rd and 4th columns of Table \ref{tab:-QF-models}. The resultant quenching factor as a function of $E_I$ is shown in Fig. \ref{fig:QF_models} in magenta.

\textbf{Sarkis $et \ al$ Model:} Another theoretical model, which is a refined form of the Linhard model, is the Sarkis \textit{$et \ al$} \cite {Sarkis:2020soy} model. This model also takes into account the internal energy of the system at the low energy recoils. We use the numerical solution of the model for germanium as given in ref. \cite {Sarkis:2020soy}
and we derive the following quenching factor as a function of $E_I$,
\begin{equation}
Q(E_I) = E_I \sum_{i=0}^{7}\left[\frac{a_i E_{I}^i}{b_i E_{I}^i}\right] \ \ \ \ (\rm Sarkis \ \textit{et al} \ \ Model)\,
\end{equation}
where $a_i$ and $b_i$ are the evaluated fit parameters and their numerical values are given in the 5th and 6th columns of Table \ref{tab:-QF-models}. The resultant quenching factor as a function of $E_I$ is shown in Fig. \ref{fig:QF_models} in green. There is about 3\% increase in the ionization efficiency in the Sarkis \textit{$et \ al$} model compared with the Linhard model as clear from the figure.

\textbf{YBe data:} Another recent direct measurement of the quenching factor for germanium is based on photo-neutron (YBe) data ref. \cite{Colaresi:2022obx} which was taken for calibrating the NCC-1701 detector to detect the CE$\overline{\nu}_e$NS of the Dresden-II reactor neutrinos. We calculate the following fitting function for the YBe data 

\begin{equation}
Q(E_I) = E_I \sum_{i=0}^{6}\left[\frac{a_i E_{I}^i}{b_i E_{I}^i}\right] \ \ \ \ (\rm YBe \ data)\,,
\end{equation}
where $a_i$ and $b_i$ are the evaluated fitting parameters. The numerical values of the fit parameters are given in the 7th and 8th columns of Table \ref{tab:-QF-models}. The related quenching factor as a function of $E_I$ is shown in Fig. \ref{fig:QF_models} in purple. Note that above 0.3 keV, no real data of Fef was used, the fit was extrapolated using the Linhard model expectations \cite{Colaresi:2022obx}.

\textbf{Jones $et \ al$ data:} We also use the old measurement of the ionization efficiency by Jones et al \cite{Jones:1975zze} because of this overlap with the ionization energy scale relevant for the nuclear recoils due to reactor neutrinos. Since the interpolation of this data results in the multi-valued quenching factor function, therefore we derive an approximate fit function for this data. The best-fit quenching factor as a function of ionization was obtained as in the following  

\begin{equation}
Q(E_I) =  \sum_{i=0}^{6}\left[\frac{E_{I}}{a_i E_{I}^i}\right] \ \ \ \ (\rm Jones \ \ \textit{et al} \ \ data)\,,
\end{equation}
where $a_i$ are the evaluated fitting parameters. The numerical values of the fit parameters are given in the 9th column of Table \ref{tab:-QF-models}. The corresponding quenching factor as a function of $E_I$ is shown in Fig. \ref{fig:QF_models} in orange.

\textbf{Fef data:} Finally, we use the iron-filtered calibration data provided with the data release file of ref. \cite{Colaresi:2022obx} and find the quenching factor as a function of the ionization energy, $E_I$, for it. The obtained fit function is given in the following
\begin{equation}
Q(E_I) = E_I \left[\frac{\sum_{i=0}^{8}a_i E_{I}^i}{\sum_{i=0}^{7} b_i E_{I}^i}\right] \ \ \ \ (\rm FeF \ data)\,,
\end{equation}
where $a_i$ and $b_i$ are the evaluated fit parameters. The numerical values of the fit parameters are given in the 11th and 12th columns of Table \ref{tab:-QF-models}. The corresponding quenching factor as a function of $E_I$ is shown in Fig. \ref{fig:QF_models} in red. There is about 15\% increase relative to the Linhard model in the range between the ionization threshold and 0.3 keV. Note that above 0.3 keV, no real data of Fef was used, the fit was extrapolated using the Linhard model expectations \cite{Colaresi:2022obx}.


\begin{table}[h!]
    \centering
    \resizebox{\columnwidth}{!}{%
   \begin{tabular}{c|c|c|c|c|c|c|c|c|c|c|c|c|}
    \multirow{3}{*}{$i$} & \multicolumn{2}{|c|}{Linhard} & \multicolumn{2}{|c|}{Bonhomme} & \multicolumn{2}{|c|}{Sarkis} & \multicolumn{2}{|c|}{YBe} & \multicolumn{2}{|c|}{Jones} & \multicolumn{2}{|c|}{Fef} \\
    & a$_i$ &  b$_i$ &  a$_i$ &  b$_i$ & a$_i$ &  b$_i$ & a$_i$ &  b$_i$ & a$_i$ &  b$_i$ & a$_i$ &  b$_i$ \\
    \hline
    0 & 0.01143   & 0.00001    & 0.00069  & 0.00001    &  0.00005   & 0.00001   &-0.60425  & 11.84055  & 1.09756   & - & 0.03200    & -0.00793     \\
    1 & 2.52766   & 0.11801    & 0.15017  & 0.00700    &  -0.0210   &-0.00050   &55.9154     &-141.433  & -4.73588    & - &-0.70192 & 0.38355      \\
    2 & 60.7988   & 19.2150    & 3.54089  & 1.11711    &  4.90730   & 0.02390   &-668.480    & 780.109  & 38.0178 & - & 6.06352    & -6.54642    \\
    3 & 248.556   & 370.305    & 14.1941  & 21.1202    &  70.4665   & 36.0068   &4017.06     &-1843.22  & -126.10   & - &-23.4161  & 55.5917       \\
    4 & 196.196   & 1246.77    & 10.9871  & 69.7824    &  84.2566   & 365.219   &-10316.7    &2011.00   & 268.504   & - & 16.2113  & -253.096      \\
    5 & 25.3242   & 804.766    & 1.39028  & 44.2150  &  12.7504    & 344.445   &12684.2     &-583.964 & -276.87   & - & 180.122    & 551.519       \\
    6 &0.13558    & 78.2333    & 0.00725  & 4.22059  &  0.21177    & 39.3241   &-5674.99    &-213.231 & 104.875   & - &-590.858    & -99.4481      \\
    7 & -     & - & -          & -        &  0.00008  & 0.39971      &  -        & -          &  -      & -         &580.171         & -1779.01      \\
    8 & -     & -              & -        & -          &   -        & -        &  -          & -       & - & -     &  -             & 2326.26      \\
\end{tabular}}%
\caption{Fit parameters for the three models. Each parameter has unit of the inverse power of keV where the powers of ‘$\rm a_i$’ correspond to the index ‘$i+1$’ and powers of ‘$\rm b_i$’ correspond to ‘$i$’.}
\label{tab:-QF-models}
\end{table}

\begin{figure}[t]
\begin{center}
\includegraphics[width=5in]{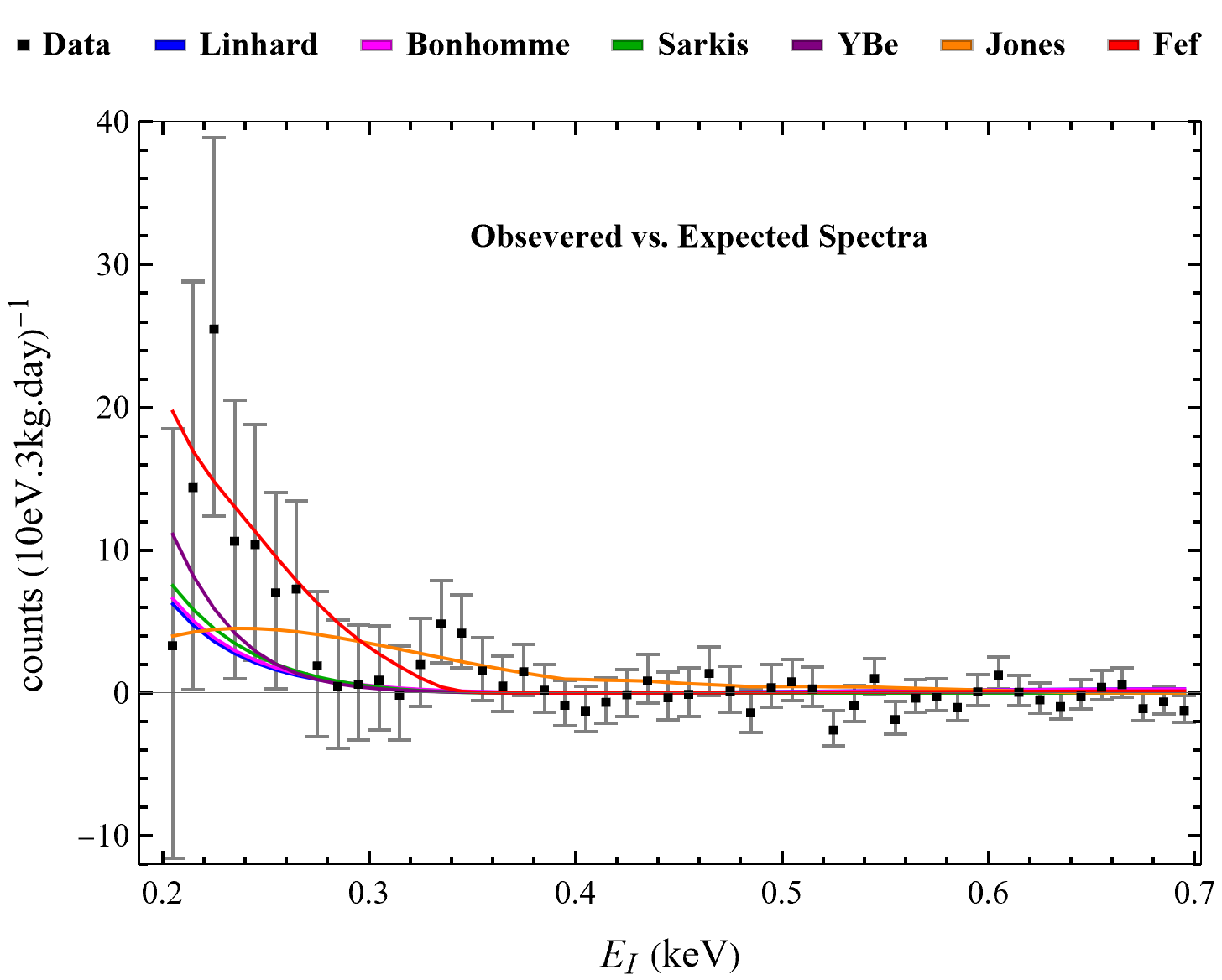}
\end{center}
\caption{Observed event energy spectrum versus predictions as a function of the true ionization energy for three quenching factors considered here.}
\label{fig:spect}
\end{figure}
Now we discuss the coherent elastic anti-neutrino nucleus scattering in the SM in addition to the form factor and introduce our notation.
\section{Coherent elastic anti-neutrino nucleus scattering}\label{sec:formalism}
At the tree level in SM, the differential cross-section 
of the reactor electron anti-neutrino scattering off the spin-0 nucleus of Germanium $\rm (^{72}Ge)$ with proton number ‘$\rm Z$’ and neutron number ‘$\rm N$’ is given by ~\cite{Freedman:1973yd, Freedman:1977xn, Tubbs:1975jx,Lindner:2016wff}, 
\begin{equation}
\frac{d\sigma_{{\overline{\nu}_e}\mathcal{N}}}{dE_{nr}}(E_{\overline{\nu}_e},E_{nr}) = \frac{G_{F}^{2}M}{\pi }
\left[Zg_{p}^{V}+ Ng_{n}^{V})\right]^{2}
\left( 1-\frac{E_{nr}}{E_{\nu}}-\frac{ME_{nr}}{2E_{{\overline{\nu}_e}}^{2}}\right) F^{2}(q^{2})\,,
\label{eq:diff-crossec}
\end{equation}%
where ‘$G_{F}$’ is the Fermi constant, ‘$E_{\overline{\nu}_e}$’ is the energy of the incoming neutrinos, 
‘$E_{nr}$’ is the nuclear recoil energy,  $q^2= -2M E_{nr}$ is the squared momentum transfer, ‘$M$’ is the mass of the target nucleus. Here, $g_{p}^{V}=(2g_{u}^{V}+ g_{d}^{V})$ and $g_{n}^{V}=(g_{u}^{V}+2g_{d}^{V})$, where $g_{u}^{V}$ and $g_{d}^{V}$ are the neutral current coupling constants for the ‘up’ and ‘down’ quarks which, in terms of the weak mixing angle ‘$\theta _{W}$’ are given by
\begin{eqnarray}
g_{u}^{V}& = & \frac{1}{2}-\frac{4}{3} \sin ^{2}\theta _{W}\,,\nonumber \\
g_{d}^{V}& = & -\frac{1}{2} +\frac{2}{3}\sin ^{2}\theta _{W}\,.
\label{eq:gv&ga}
\end{eqnarray}%
To include all the radiative corrections, we will use $\rm sin ^{2}\theta_{W}=0.23857 \pm 0.00005$, the low energy value evaluated in $\overline{\rm MS}$ scheme \cite{Marciano:1980pb,Erler:2004in,ParticleDataGroup:2020ssz}.
In eq. (\ref{eq:diff-crossec}), $F(q^{2})$ is nuclear form factor, where  we use the Klein-Nystrand form factor \cite{Klein:1999gv} as given in the following
\begin{equation}
F(q^{2})=\frac{4\pi \rho _{0}}{Aq^{3}}[\sin (qR_{A})-qR_{A}\cos (qR_{A})]%
\left[ \frac{1}{1+a^{2}q^{2}}\right],  \label{F-bessel}
\end{equation}%
where $\rho _{0}$ is the normalized nuclear number density, $A$ is the  atomic number of $\rm ^{72}Ge$, $%
R_{A}=1.2A^{1/3}\, \mathrm{fm}$ is the nuclear radius, and $a=0.7\, \mathrm{fm
 }$ is the range of the Yukawa potential. At such low energy recoils, the form factor effects are less than $3\%$.
\section{Predicted spectrum and the statistical analysis}\label{sec:analysis}
The Germanium based NCC-170 detector of total mass 3 kg receives electron anti-neutrinos from Dresden-II boiling water reactor of thermal power of 2.96 $\rm GW_{th}$. The detector is located at distance $d=8$ m from the reactor source. For this experimental setup, the total number of events of the nuclear recoil in a given bin ‘$i$’ of the ionization energy reads,
\begin{equation} 
N^i=n.t. 
\int_{E_{I}^{\prime \rm i}}^{E_{I}^{\prime \rm i+1}}dE_{I}^{\prime}
\int_{0}^{E_I^{\rm max}}dE_{I} 
\int_{E_{\overline{\nu}_e}^{\rm min }}^{E_{\overline{\nu}_e}^{\rm max}}dE_{\overline{\nu}_e}
\frac{d\sigma_{{\overline{\nu}_e}\mathcal{N}}}{dE_{I}}(E_{\nu },E_{I})\frac{%
d\phi _{\overline{\nu}_e}({E_{\overline{\nu}_e} })}{dE_{\nu }} G( E_{I}^\prime, E_{I}) \mathcal{E}\\ (E^\prime), 
\label{eq:eventrt}
\end{equation}%
where $n =2.43 \times 10^{25}$ is the number of target nuclei corresponding to the 2.924 kg fiducial mass of $\rm ^{72}Ge$, t = 96.4 days is the data taking time,  $E_{I}$ denotes the actual ionization energy and $E_{I}^\prime$ denotes the visible ionization energy. The integration range of the $E_{I}^\prime$ was taken 10 eV according to the observed spectrum. Here, $E_I^{\rm max}=
2E_{\overline{\nu}_e}^{2}/(2E_{\overline{\nu}_e}+M)
$ is maximum true ionization energy, $E_{\overline{\nu}_e}^{\rm min } = E_{nr} + \sqrt{E_{nr}^2+2M E_{nr}}/2$ is the minimum anti-neutrino energy that produces a nuclear recoil, $E_{nr}$,  while the $E_{\overline{\nu}_e}^{\rm max}$ is the endpoint of the reactor neutrino spectrum, which we take as 9 MeV. Note that $E_{\overline{\nu}_e}^{\rm min }$ is expressed as a function of $E_I$ using eq. (\ref{eq:QF}). Here, $\frac{d\sigma_{{\overline{\nu}_e}\mathcal{N}}}{dE_{I}}$ is the differential cross-section as a function of the true ionization energy that can be obtained from eq. (\ref{eq:QF}) and (\ref{eq:diff-crossec}) using the derivative chain rule. The result can be written as,

\begin{equation}
\frac{d\sigma_{{\overline{\nu}_e}\mathcal{N}}}{dE_{I}}(E_{\nu},E_{I}) = \left(\frac{Q(E_I)- E_I\frac{dQ(E_I)}{dE_I}}{Q^2(E_I)}\right) \left(\frac{d\sigma_{{\overline{\nu}_e}\mathcal{N}}}{dE_{nr}}(E_{\nu},E_{nr})
\Bigr|_{\substack{E_{nr}= \frac{E_I}{Q(E_I)}}}\right).
\label{eq:diff-crossecion}
\end{equation}%
The reactor anti-neutrino energy spectrum ($\frac{d\phi _{\overline{\nu}_e}}{{dE_{\nu }}}(E_{\nu })$) for energies above $\sim$ 2 MeV is given by \cite{Kopeikin:2004cn},
\begin{equation}
\frac{d\phi _{\overline{\nu}_e}}{{dE_{\nu }}}(E_{\nu }) = \frac{n_{f}}{4\,\pi\, d^{2}}\left(\sum_i f_i \exp{\left[a_{0i}+ \left(\sum_{j=1}^5 a_j E_{\nu}^j\right)_i\right]} \right)\,\text, 
\end{equation}
where $n_f= 9 \times 10^{19}$ fissions per second corresponding to  2.96 Giga-watt thermal power and an average energy release of 205.25 MeV by all reactor components, ‘$d$’ is the distance of the detector from the reactor core and ‘$i$’ sums over the fuel constituents ${^{235}\mathrm{U}}, {^{238}\mathrm{U}}\ {^{239}\mathrm{Pu}}, {^{241}\mathrm{Pu}}$. The coefficients, $f_i$, which represent the fission rate of each component element and the coefficients of the neutrino energy $E_{\nu}$ in the exponent (a$_0$ and a$_j$) were both taken from ~\cite{Mueller:2011nm}. The low energy part of the flux spectrum ($\lesssim$ 2MeV) is mainly governed by the slow neutron capture by the ${^{238}\mathrm{U}}$. We use the numerical data for this part of the spectrum, taken from ref. \cite{TEXONO:2006xds}.

$G(T^\prime, T)$ is the Gaussian distribution function which accounts for the detector energy resolution as given in the following,
\begin{equation}
G(E^{\prime}_I,E_I)=\frac{1}{\sqrt{2\pi\sigma^2}}\exp{\left[-\frac{(E_I^{\prime}-E_I)^2}{2\sigma^2}\right]}\,.
\label{eq:smearing}
\end{equation}%
Here, the Gaussian width ‘$\sigma$’ is given by $\sigma= \sqrt{\sigma_n^2+E_{I}\eta F}$ where $\sigma_n=68.5$~eV is the electronic noise, $\eta=2.96$~eV is the average energy of photons to create an electron-hole pair in germanium, and $F= 0.105$ is the Fano factor \cite{Colaresi:2022obx}. Finally, $\mathcal{E} (E^\prime)$ represents the signal acceptance. The measured data has already been corrected for the signal acceptance \cite{Colaresi:2022obx}, therefore, we do not use it in calculating our predicted spectrum. Note that we normalize the integrand of the variable $E_I$ in eq. (\ref{eq:eventrt}) with the integral of the Gaussian function of eq. (\ref{eq:smearing}).

To fit our expected spectrum for $\sin^2{\theta_W}$ and electromagnetic interaction parameters to the observed data, we use the following $\chi^2$ function,
\begin{equation}
\chi ^{2}=\underset{i=1}{\overset{50}{\sum }}
\left(\frac{N_{\rm obs}^{i}-N_{\rm exp}^{i}(1+\alpha)}
{\sigma^{i}}\right)^2
+\left( \frac{\alpha }{\sigma _{\alpha }}\right) ^{2} \,,
\label{eq:chisq}
\end{equation}%
where $N_{\rm obs}^{i}$ is the background-subtracted observed events in the $i-$th energy bin given in units of per 10 eV per 3 kg per day as shown in Fig. \ref{fig:spect}, which was taken from \cite{Collar:2022ibv}, $N_{\rm exp}^{i}$ is the expected events in the corresponding energy bin, $\sigma_i$ is the uncertainty in the data points as shown in error bars in Fig. \ref{fig:spect}, which include the combination of the signal acceptance and statistical uncertainties. The pull term in eq. (\ref{eq:chisq}) is added in to account for the theoretical uncertainties, ‘$\alpha$’ is the pull parameter and $\sigma_{\alpha} = \sqrt{\sigma_{f}^2 + \sigma_{qf}^2}$ is the total theoretical uncertainty, where $\sigma_{f}= 5\%$ is reactor flux total uncertainty and $\sigma_{qf}$ is the uncertainty in each QF considered here. The average uncertainty in the Sarkis $et \ al$ model is $\sigma_{qf}= 25\%$ uncertainty. We use the same value also for the Linhard model. On the other hand, we use the average value for the uncertainty on Fef data, which is $\sigma_{qf}= 40\%$ as shown in Fig. (4) of ref. \cite{Colaresi:2022obx} and likewise for the other quenching factors. 

Next, we present and discuss our results. We will analyze the observed data for the $\sin^2{\theta_W}$, neutrino millicharge, magnetic moment, charge radius and neutrino anapole moment.

\section{Weak mixing angle at keV nuclear recoils}\label{sec:EW}
\begin{figure}[t]
\begin{center}
\includegraphics[width=5.0in]{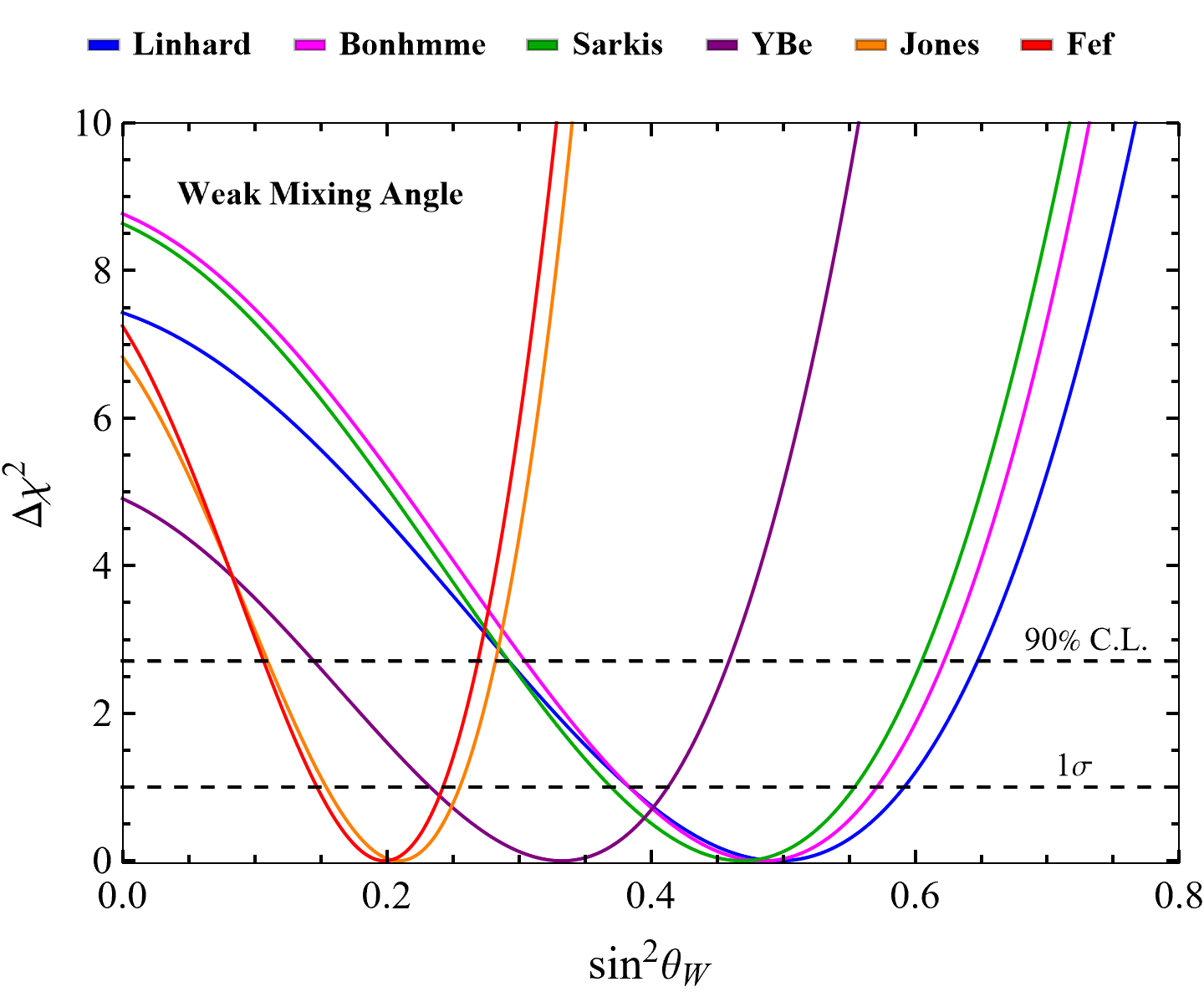}
\label{fig:wma}
\end{center}
\caption{$\Delta \chi^2$ distribution of $\sin^2{\theta_W}$ with 1$\sigma$ and $90\%$ C.L. projections using two theoretical models and the iron-filtered data for the quenching factors. See text for discussion.}
\label{fig:EW-angle}
\end{figure}

With the statistical model introduced in the preceding section, we first discuss the implications of the observed reactor neutrino coherent scattering process for the weak mixing angle, namely $\sin^2{\theta_W}$, using the three different quenching factors discussed in Sec. \ref{sec:QF}. We fit $\sin^2{\theta_W}$ using the two theoretical models and the ion-filter data for the QF. The results of the parameter fitting are shown in Fig. \ref{fig:EW-angle} in the form of 1-dimensional $\Delta \chi^2$ distributions. The best-fit values with 1 $\sigma$ uncertainties for the three cases are given in the following,
\begin{alignat}{2}
  \label{weak_mix_angle_1_sigma}
  \quad&\sin^2\theta_W=0.50^{+0.09}_{-0.12}& \ \ \ \ (\text{Lindhard Model}),\ \nonumber\\
   \quad&\sin^2\theta_W=0.48^{+0.14}_{-0.18}& \ \ \ \ (\text{Bonhomme \textit{et al} data}),\ \nonumber\\
  \quad&\sin^2\theta_W=0.47^{+0.08}_{-0.11}& \ \ \ \ (\text{Sarkis $et \ al$ Model}),\ \nonumber\\
  \quad&\sin^2\theta_W=0.33^{+0.13}_{-0.18}& \ \ \ \     (\text{YBe data}), \ \nonumber\\
  \quad&\sin^2\theta_W=0.22^{+0.06}_{-0.11}& \ \ \ \     (\text{Jones data}), \ \nonumber\\
  \quad&\sin^2\theta_W=0.20^{+0.04}_{-0.05}& \ \ \ \     (\text{Fef data}).\
\end{alignat}

By comparing the three results, it is clear how sensitive the best-fit values and the corresponding uncertainties are to QF. The theoretically predicted value at the related energy scale in the $\overline{\rm MS}$ scheme is $\rm sin ^{2}\theta_{W}=0.23857$,  \cite{Marciano:1980pb,Erler:2004in,ParticleDataGroup:2020ssz}. This gives, respectively, 52\%, 49\% and 14\% discrepancy with the Linhard model, Sarkis $et \ al$ model and with the Fef data. The rest accure in this range. However, notice that the percentage change in the case of two models is increasing while in the case of Fef data it is decreasing. This implies that the theoretical models for the QF overestimate the true value of $\rm sin ^{2}\theta_{W}$ while the Fef data underestimate its value, therefore, the true value lies in between the two types of extreme values of the fits. Notice that the Sarkis $et \ al$ model improves the agreement by 3\%.

This is the first determination of the $\rm sin ^{2}\theta_{W}$ at the sub-keV nuclear recoils using the CE$\overline{{\nu}_e}$NS data with the reactor anti-neutrinos. Before it was determined using the COHERENT data \cite{Akimov:2017ade, Akimov:2018vzs,Akimov:2021dab} with nuclear recoils above $\sim $ 5 keV \cite{Khan:2019cvi, Cadeddu:2019eta, Cadeddu:2020lky}. The results obtained here show that we need a better understanding for the quenching factor. More importantly, the new data for the QF needs more precision. Also, its consistency with theoretical modeling is essential.

\section{Electromagnetic interactions of neutrinos at keV nuclear recoils}\label{sec:EMprorties}
\subsection{Millicharge neutrinos}

Electric charge quantization is assumed in the standard model, however, several theories beyond the SM like those with magnetic monopoles \cite{Dirac:1931kp}, grand unified theories \cite{Georgi:1974sy,Pati:1974yy} and the extra dimension models \cite{Arkani-Hamed:2006emk} predict the charge quantization. Other extensions predict new particles with fractional charges \cite{Ignatiev:1978xj,Okun:1983vw,Holdom:1986eq,Kors:2004dx,Batell:2005wa}, which can be promising candidates for dark matter \cite{Goldberg:1986nk, Mohapatra:1990vq, Kors:2005uz,Gies:2006ca,Cheung:2007ut, Feldman:2007wj,Berlin:2019uco, Berlin:2021kcm}. Among the SM particles, neutrinos are the most promising candidates for such particles, often called milli-charged neutrinos \cite{Babu:1989tq,Babu:1989ex,Foot:1990uf,Foot:1992ui}. The electric charge dequantization is also related to the emergence of the gaugeable $U(1)$ symmetries, $L_e-L_{\mu}$, $L_{\mu}-L_\tau$ and $L_e-L_{\tau}$ and $B-L$. Only one of the three differences can be anomaly-free and the corresponding difference is added to the hypercharge of the SM which leads to the fractional charges of Dirac-type neutrinos \cite{Foot:1990uf,Foot:1992ui, Babu:1989tq}. In these simple extensions of the SM, the value of neutrino millicharges is arbitrary and only experimental \cite{Davidson:1991si, Babu:1993yh,Bressi:2011yfa, Gninenko:2006fi,Chen:2014dsa,TEXONO:2018nir,Khan:2019cvi,Cadeddu:2019eta,Cadeddu:2020lky,Khan:2020vaf} or observational \cite{Barbiellini:1987zz, Raffelt:1999gv, Davidson:2000hf,Melchiorri:2007sq,Studenikin:2012vi} limits are available. 

The contribution of the neutrino millicharges to the SM weak interaction process of the coherent neutrino-nucleus ($\overline{\nu}-N$) scattering is parameterized in terms of $Q_{\overline{\nu}_\alpha}$ and the interaction term is given by
\begin{equation}
\mathcal{L}_{\alpha}^{em}=-ie
\left(Q_{\overline{\nu}_\alpha}{\overline{\nu}_{\alpha}}\gamma_{\mu }{\nu}_{\alpha} + {\overline{N}}\gamma_{\mu}{N} \right) A^{\mu},
\label{eq:CHargN}
\end{equation}
where $A^{\mu}$ is mediating electromagnetic field and ‘$e$’ is the unit electric charge. In principle, neutrinos with intrinsic electric charge should be negatively charged while the antineutrinos should be positively charged, however, since this is a direct way of probing the property, we, therefore, make no distinction in the sign of neutrino and anti-neutrinos. Thus, we add its contribution to the SM interactions. Further, the electromagnetic interactions due to the intrinsic electric charge of neutrinos add up coherently to the vector part of the SM weak interaction, therefore, its effect on reactor anti-neutrinos can be included through the weak mixing angle in eq. (\ref{eq:gv&ga}) accordingly as,
\begin{eqnarray}
\sin ^{2} \theta _{W}\rightarrow \sin ^{2}\theta _{W} \left(1 - \frac{\pi \alpha_{em}}{\sqrt{2} \sin ^{2}\theta _{W} G_F ME_{nr}}Q_{{\overline{\nu}_e}}\right),
\label{eq:milicharge}
\end{eqnarray}
where $\alpha_{em}$ is the fine structure constant.

\begin{figure}[t]
\begin{center}
\includegraphics[width=5.0 in]{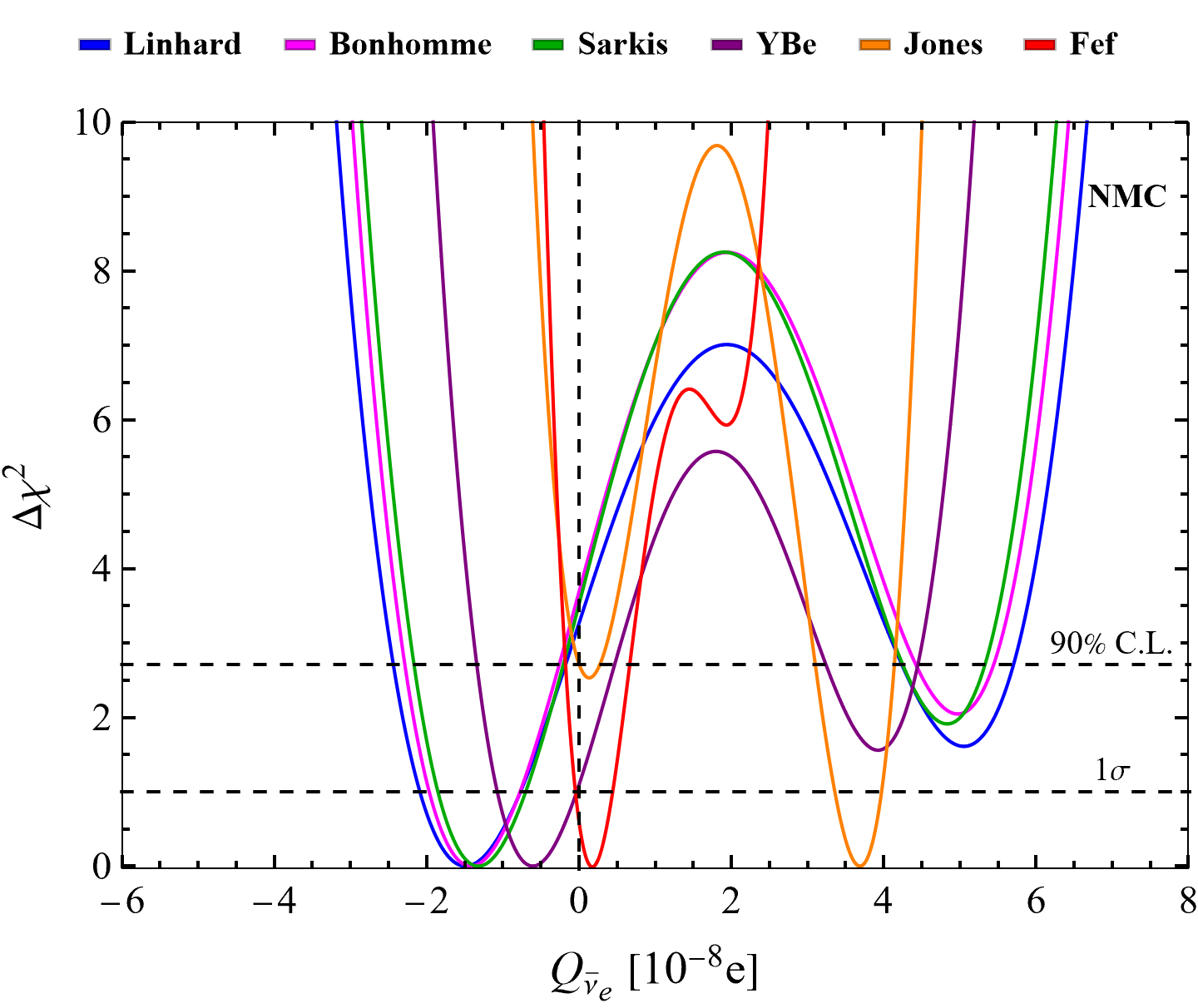}
\end{center}
\caption{$\Delta \chi^2$ distributions of neutrino millicharge (NMC) with 1$\sigma$ and $90\%$ C.L. projections using two theoretical models and the iron-filtered data for the quenching factors. See text for details.}
\label{fig:NMC}
\end{figure}
Notice that for the CE$\overline{\nu}_e$NS with reactor anti-neutrinos there is only one parameter involved, namely, $Q_{\overline{\nu}_e}$. We determine the constraints on neutrino millicharge by fitting, $Q_{\overline{\nu}_e}$ using the two theoretical models for the QF and the Fef data. The results of the parameter fitting are shown in Fig. \ref{fig:NMC} in the form of 1-dimensional $\Delta \chi^2$ distributions. At 90\% C.L., we obtain the following constraints,
\begin{empheq}[box=\widefbox]{align}
-2.43\times 10^{-8}<Q_{\overline{\nu}_e}/e<5.73\times 10^{-8}\, \ \ \ \ \ \ \ \ \ (\text{Lindhard Model})\  \nonumber \\
-2.26\times 10^{-8}<Q_{\overline{\nu}_e}/e<5.45\times 10^{-8}\,  \ \ (\text{Bonhomme \textit{et al} data})\  \nonumber \\
-2.11\times 10^{-8}<Q_{\overline{\nu}_e}/e<5.32\times 10^{-8}\, \ \ \ \ \ \ (\text{Sarkis \textit{et al} Model})\  \nonumber \\
-1.36\times 10^{-8}<Q_{\overline{\nu}_e}/e<4.46\times 10^{-8}\, \ \ \ \ \ \ \ \ \ \ \ \ \ \ \ \ \ \ (\text{YBe data})\  \nonumber \\
-0.09\times 10^{-8}<Q_{\overline{\nu}_e}/e<4.15\times 10^{-8}\, \ \ \ \ \ \ \ \ \ \ \ \ \ \ \ \ (\text{Jones data})\  \nonumber \\
-0.16\times 10^{-8}<Q_{\overline{\nu}_e}/e<0.66\times 10^{-8}\, \ \ \ \ \ \ \ \ \ \ \ \ \ \ \ \ \ \ \ (\text{Fef data})\
\end{empheq}

In the case of Linhard and Sarkis \textit{et al} models, the bounds are comparable to those obtained before from the coherent scattering using COHERENT data \cite{Khan:2022not} while from the Fef QF data they are two orders of magnitude stronger than the previous bounds \cite{Khan:2022not}.

It is important to note that for the three quenching factors considered here, there is a mild preference for the non-zero neutrino millicharges as shown in Fig. \ref{fig:NMC}. 
This preference can be attributed to the interference between the SM and the neutrino millicharge term
and its dependence on the inverse double power of the nuclear recoil energy and on the target mass in the cross-section (see eq. (\ref{eq:milicharge}) in combination to eq. (\ref{eq:diff-crossec})). This is unlike the other electromagnetic interactions. These aspects of the coherent scattering process have been discussed in detail in ref. \cite{Khan:2022not}.

Stronger limits on millicharge neutrinos are available from the observational studies \cite{Barbiellini:1987zz, Raffelt:1999gv, Davidson:2000hf,Melchiorri:2007sq,Studenikin:2012vi}. The strongest available upper limit is $Q_\nu \leq 2 \times 10^{-15} e $ which comes from the time arrival dispersion and the energy spread of neutrinos from SN1987A \cite{Barbiellini:1987zz}. The laboratory bounds from the $\nu-e$ are also several orders of magnitude from the limits derived here \cite{Davidson:1991si, Babu:1993yh,Bressi:2011yfa, Gninenko:2006fi,Chen:2014dsa,TEXONO:2018nir,Khan:2019cvi,Cadeddu:2019eta,Cadeddu:2020lky,Khan:2020vaf}. For instance, the TEXONO experiment derives the limit, $Q_\nu \leq 2.1 \times 10^{-12} e $. However, this difference can be easily understood from the kinematical considerations \cite{Khan:2022not}. It was shown in ref. \cite{Khan:2022not} that robustness of the bounds or preference of non-zero neutrino millicharges totally depends on the experimental precision.

\subsection{Neutrino magnetic moment}
The neutrino magnetic moment in the coherent neutrino-nucleus scattering process has been studied before and constraints with the COHERENT data were derived before in ref. \cite{Khan:2019cvi, Cadeddu:2019eta}. In the general coupling for Majorana ($M$) or Dirac ($D$) neutrinos to the electromagnetic field strength ($F^{\mu\nu}$), the neutrino magnetic moment interaction term can be written as \citep{Fujikawa:1980yx, Shrock:1982sc, Vogel:1989iv,Abak:1989kp, Grimus:1997aa}
\begin{equation}
{\cal L}^M = -\frac 14 \bar \nu_{\alpha L}^c \, \lambda_{\alpha \beta}^M \, \sigma_{\mu\nu} \, \nu_{\beta L} \, F^{\mu\nu}~\mbox{ or }~{\cal L}^D =-\frac 12 \bar \nu_{\alpha R} \, \lambda_{\alpha \beta}^D \, \sigma_{\mu\nu} \, \nu_{\beta L}\, F^{\mu\nu}\,,
\label{eq:MM}
\end{equation}
where $\lambda^X = \mu^X - i \epsilon^X$, which is hermitian for the Dirac neutrinos and antisymmetric for Majorana neutrinos. For Majorana neutrinos, only transition magnetic moments are possible while the flavor diagonal is zero. Here, we consider the flavor diagonal neutrino magnetic moment of reactor electron antineutrino ($\mu_{\overline{\nu}_e}$). Again, there is only one parameter involved. The SM prediction of the non-zero neutrino magnetic moment at a loop level can be quantified as below, \citep{Fujikawa:1980yx, Vogel:1989iv} 
\begin{equation}
\mu_{\overline{\nu}_e} = \frac{3eG_F \rm m_{\overline{\nu}_e}}{8 \sqrt{2} \pi^2} \sim 3 \times 10^{-19} \mu_{B}\left(\frac{\rm m_{\overline{\nu}_e}}{1 \rm eV}\right).
\end{equation}


As clear from eq. (\ref{eq:MM}), for Dirac neutrinos, the helicity of the final state neutrino changes in interaction due to its magnetic moment, therefore, no interference with the SM weak interaction can occur. The corresponding contribution adds to the SM weak cross-section incoherently. We can write down the differential cross-section for the neutrino magnetic moment (MM) of electron anti-neutrinos scattering off a spin-0 nucleus of $^{72}\rm Ge$ with proton number (Z) as in the following \cite{Khan:2019cvi},
\begin{equation}
\frac{d\sigma_{{\overline{\nu}_e}\mathcal{N}}^{MM}}{dE_{nr}}(E_{\overline{\nu}_e},E_{nr}) =\left(\frac{\pi \alpha_{em}^2\,\mu _{\overline{\nu}_e}^2}{m_{e}^2}\right)%
\text{ }\left(\frac{1}{E_{nr}}-\frac{1}{E_{\overline{\nu}_e}}+\frac{E_{nr}}{4E_{\overline{\nu}_e} ^{2}}\right) Z^2 F^{2}(q^{2}), 
\label{eq:EDM}
\end{equation}
where $\mu_{\overline{\nu}_e}$ is expressed in units of Bohr's magneton ($\mu_B$) and $m_e$ is the electron mass. One can notice that in comparison to the millicharge neutrinos, as given in eq. (\ref{eq:milicharge}) in combination with eq. (\ref{eq:diff-crossec}), the neutrino magnetic moment has no interference with the SM weak interaction and the dependence on the inverse power of the nuclear recoil is only linear in the leading term. Notice that eq. (\ref{eq:EDM}) can be written in terms of the ionization energy $E_I$ in a similar fashion as described for the SM cross-section around eq. (\ref{eq:diff-crossecion}). 

\begin{figure}[t]
\begin{center}
\includegraphics[width=5.0 in]{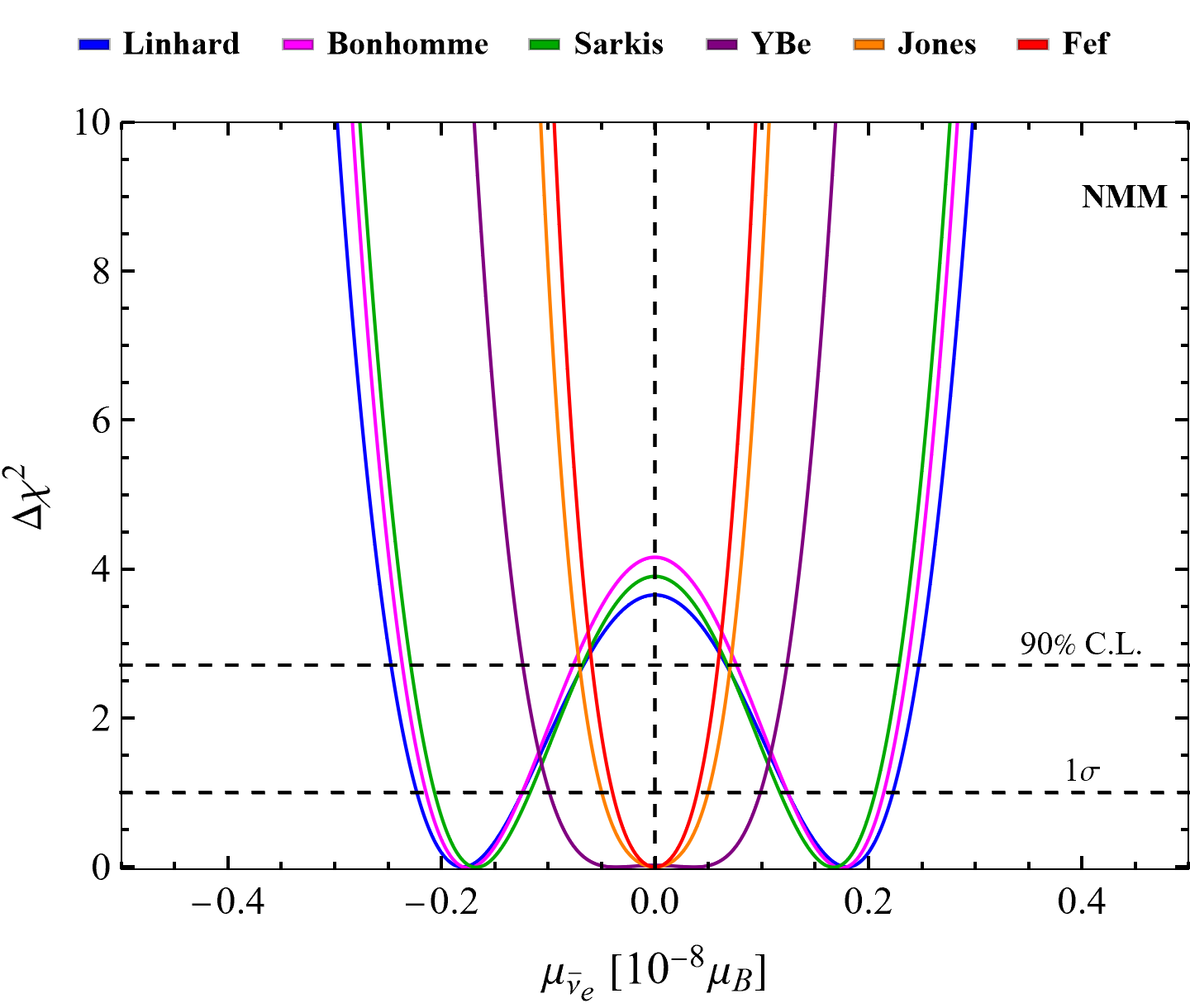}
\end{center}
\caption{$\Delta \chi^2$ distributions of neutrino magnetic moment (NMM) with 1$\sigma$ and $90\%$ C.L. projections using two theoretical models and the iron-filtered data for the quenching factors. See text for details.}
\label{fig:NMM}
\end{figure}

Using eq. (\ref{eq:EDM}) in terms of the ionization energy in addition to eq. (\ref{eq:diff-crossecion}) and our $\chi^2$ function defined in eq. (\ref{eq:chisq}), we fit $\mu_{\overline{\nu}_e}$ to the data and derive the constraints according to the three QFs. The results are shown in Fig. \ref{fig:NMM} in the form $\Delta \chi^2$ profile. For guidance, we also show the 1$\sigma$ and 90\% C.L. projections in the figure. We obtain the following bounds at 90\% C.L.,  

\begin{empheq}[box=\widefbox]{align}
-0.25\times 10^{-8}<\mu_{\overline{\nu}_e}/\mu_B<0.25\times 10^{-8}\, \ \ \ \ \ \ \ \ \ \ (\text{Lindhard Model})\  \nonumber \\
-0.24\times 10^{-8}<\mu_{\overline{\nu}_e}/\mu_B<0.24\times 10^{-8}\,  \ \ (\text{Bonhomme \textit{et al} data})\  \nonumber \\
-0.23\times 10^{-8}<\mu_{\overline{\nu}_e}/\mu_B<0.23\times 10^{-8}\, \ \ \ \ \ \ (\text{Sarkis \textit{et al} Model})\  \nonumber \\
-0.12\times 10^{-8}<\mu_{\overline{\nu}_e}/\mu_B<0.12\times 10^{-8}\, \ \ \ \ \ \ \ \ \ \ \ \ \ \ \ \ \ \ (\text{YBe data})\  \nonumber \\
-0.07\times 10^{-8}<\mu_{\overline{\nu}_e}/\mu_B<0.07\times 10^{-8}\, \ \ \ \ \ \ \ \ \ \ \ \ \ \ \ \ (\text{Jones data})\  \nonumber \\
-0.06\times 10^{-8}<\mu_{\overline{\nu}_e}/\mu_B<0.06\times 10^{-8}\, \ \ \ \ \ \ \ \ \ \ \ \ \ \ \ \ \ \ \ (\text{Fef data})\
\end{empheq}
In the case of Linhard and Sarkis \textit{et al} models, the bounds are comparable to those obtained before from the coherent scattering using COHERENT data \cite{Khan:2022not} while from the Fef QF data they are a factor of seven stronger than the previous bounds \cite{Khan:2022not}.

\subsection{Neutrino charge radius}
\begin{figure}[t]
\begin{center}
\includegraphics[width=5.0in]{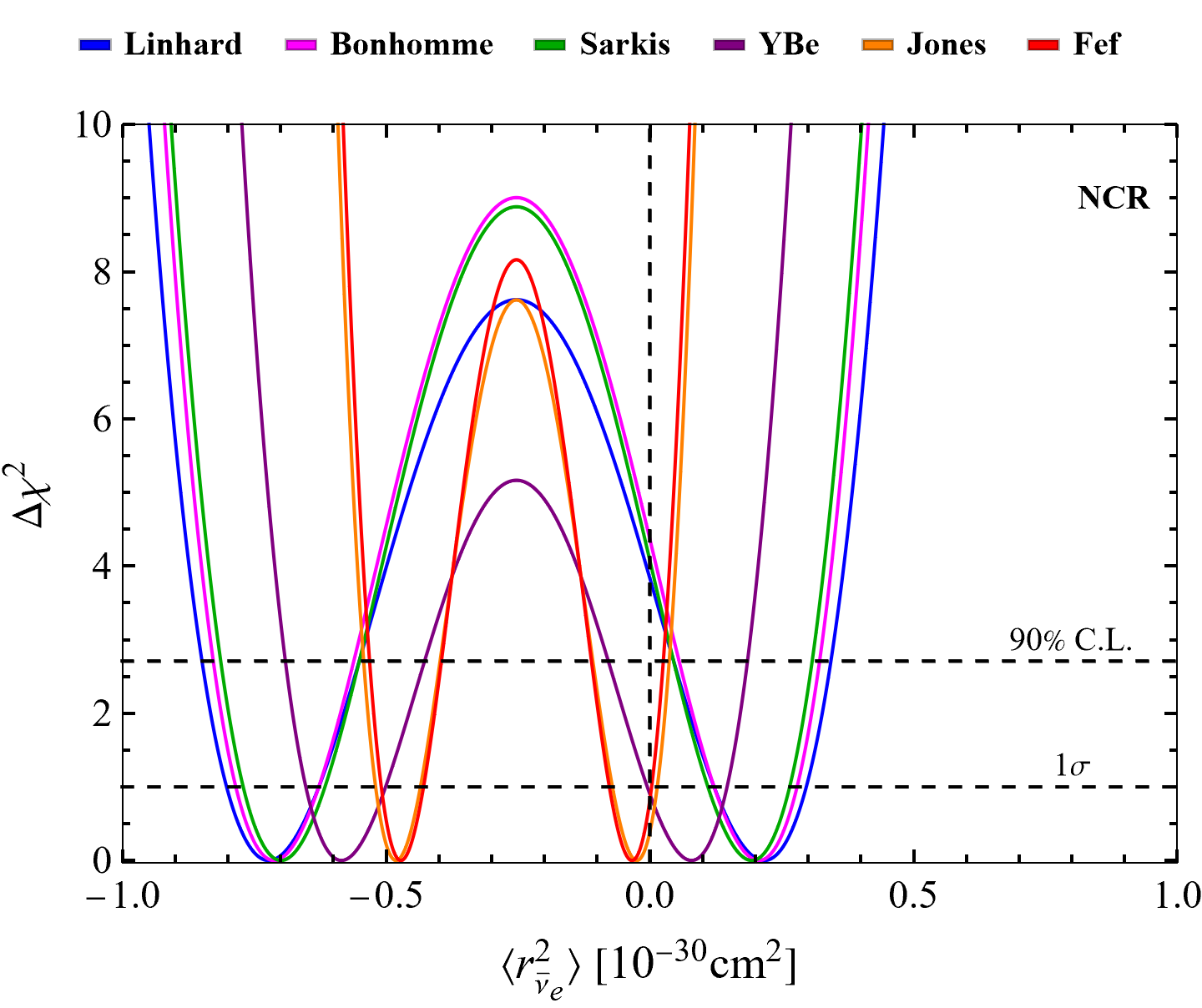}
\end{center}
\caption{$\Delta \chi^2$ distributions of neutrino charge radius (NCR) with 1$\sigma$ and $90\%$ C.L. projections using two theoretical models and the iron-filtered data for the quenching factors. See text for details.}
\label{fig:NCR}
\end{figure}

In the SM, the neutrino charge radius for neutrinos is induced by radiative corrections. 
Its relevance to CE$\nu$NS has been discussed before in ref. \cite{Papavassiliou:2005cs}. For the general effective electromagnetic vertex of massive neutrinos, $\bar \nu \Lambda_\mu \nu A^\mu$, the neutrino charge radius term is written as \cite{Bernabeu:2000hf,Bernabeu:2002nw,Bernabeu:2002pd,Fujikawa:2003ww,},
\begin{equation}
\Lambda _{\mu }(q)=\gamma _{\mu }F_{\nu}(q^{2}) \simeq \gamma 
_{\mu }q^{2}\frac{\langle r^{2}\rangle }{6}\,,
\end{equation}
where $q$ is the momentum transfer and $F_{\nu}(q^{2}) $ is a form 
factor that is related to the neutrino charge radius $\langle r_{\nu}^{2}\rangle $ via 
\begin{equation}
\langle r_{\nu}^{2}\rangle =6\left. \frac{dF_{\nu }(q^{2})}{dq^{2}} \right|_{q^{2}=0} \,.
\label{eq:ncrs}
\end{equation}
Notice that choice for the sign in the definition of the charge radius in eq. (\ref{eq:ncrs}) is completely conventional. Here, we consider positive signs. The SM prediction of the neutrino charge radius, therefore, is \cite{Bernabeu:2000hf,Bernabeu:2002nw,Bernabeu:2002pd,Fujikawa:2003ww,Novales-Sanchez:2013rav,Cadeddu:2Sarkis018dux},
\begin{equation}
\langle r_{\nu_\alpha }^{2}\rangle _{\rm SM}=-\frac{G_{F}}{2\sqrt{2}\pi }\left[
3-2\ln \left(\frac{m_{\alpha}^{2}}{m_{W}^{2}}\right) \right] ,
\end{equation}
where $m_{\alpha}$ is the mass of the charged lepton associated to $\nu_{\alpha}$ and $m_W$ is the mass of the $W^{\pm}$ boson. The numerical values for the electronic flavor in the SM therefore is \cite{Bernabeu:2000hf,Bernabeu:2002nw,Bernabeu:2002pd,Novales-Sanchez:2013rav,Cadeddu:2018dux} 
\begin{eqnarray}
\langle r_{\overline{\nu}_e}^{2}\rangle _{\rm SM}=-0.83 \times 10^{-32} \ \ \rm cm^{2}.
\end{eqnarray}
Like the neutrino millicharges, the neutrino charge radii contribute coherently to the SM process \cite{Papavassiliou:2005cs} and its effect on the CE$\overline{\nu}$NS process can be added to the weak mixing angle by making the following replacement in eq. (\ref{eq:gv&ga}), 
\begin{eqnarray}
\sin ^{2} \theta _{W}\rightarrow \sin ^{2}\theta _{W} \left(1 + \frac{\pi \alpha_{em}}{3 \sqrt{2} \sin ^{2}\theta _{W} G_F} \langle r_{\nu_\alpha }^{2}\rangle \right).
\label{eq:NCR}
\end{eqnarray}
We note that, unlike the millicharge neutrinos in eq. (\ref{eq:milicharge}), the charge radius does not have dependence on the inverse power of the recoil energy and on the target mass. Therefore, a weaker sensitivity compared to the millicharge neutrinos is expected. This was also noted before in 
refs \cite{Khan:2020vaf, Khan:2017djo} and recently for analysis with the COHERENT data \cite{Khan:2022not}. Here, the only parameter that contribute is the $\langle r_{\overline{\nu}_e }^{2}\rangle$ and we fit this parameter using the observed data of the sub-keV nuclear recoils. The obtained results for the three QFs are shown in Fig. \ref{fig:NCR} and the constraints obtained at 90\% C.L. are the following,

\begin{empheq}[box=\widefbox]{align}
-0.85\times 10^{-30}<\langle r^2_{\overline{\nu}_e}\rangle/\rm cm^2<0.35\times 10^{-30}\,  \ \ \ \ \ \ \ \ \ \ (\text{Lindhard Model})\  \nonumber \\
-0.82\times 10^{-30}<\langle r^2_{\overline{\nu}_e}\rangle/\rm cm^2<0.32\times 10^{-30}\,  \ \ (\text{Bonhomme \textit{et al} data})\  \nonumber \\
-0.80\times 10^{-30}<\langle r^2_{\overline{\nu}_e}\rangle/\rm cm^2<0.30\times 10^{-30}\, \ \ \ \ \ \ (\text{Sarkis \textit{et al} Model})\  \nonumber \\
-0.70\times 10^{-30}<\langle r^2_{\overline{\nu}_e}\rangle/\rm cm^2<0.18\times 10^{-30}\, \ \ \ \ \ \ \ \ \ \ \ \ \ \ \ \ \ \ (\text{YBe data})\  \nonumber \\
-0.55\times 10^{-30}<\langle r^2_{\overline{\nu}_e}\rangle/\rm cm^2<0.04\times 10^{-30}\, \ \ \ \ \ \ \ \ \ \ \ \ \ \ \ \ (\text{Jones data})\  \nonumber \\
-0.50\times 10^{-30}<\langle r^2_{\overline{\nu}_e}\rangle/\rm cm^2<0.03\times 10^{-30}\, \ \ \ \ \ \ \ \ \ \ \ \ \ \ \ \ \ \ \ (\text{Fef data})\
\end{empheq}
In the case of Linhard and Sarkis \textit{et al} models, the bounds are comparable to those obtained before from the coherent scattering using COHERENT data \cite{Khan:2022not} while from the Fef QF data are a factor of five stronger than the previous bounds \cite{Khan:2022not}.

\subsection{Neutrino anapole moment}
\begin{figure}[t]
\begin{center}
\includegraphics[width=5.0in]{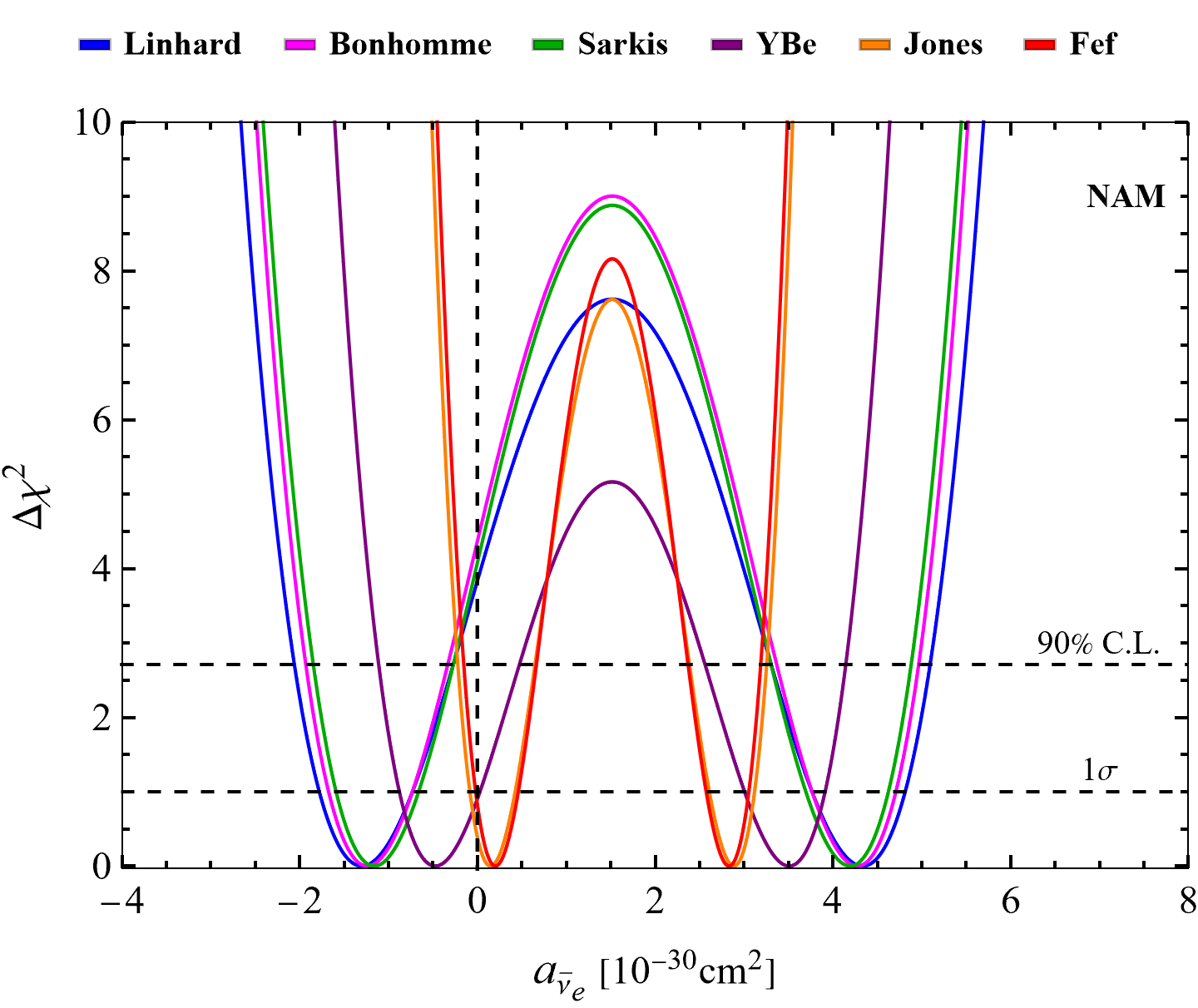}
\end{center}
\caption{$\Delta \chi^2$ distributions of neutrino anapole moment (NAM) with 1$\sigma$ and $90\%$ C.L. projections using two theoretical models and the iron-filtered data for the quenching factors. See text for details.}
\label{fig:NAM}
\end{figure}

If neutrino carries a non-zero charge radius, it can also have a non-zero anapole moment which is induced from the gauge-invariant parity-odd axial current \cite{zel1958electromagnetic, zel1960effect, Barroso:1984re, Abak:1987nh, Musolf:1990sa,Dubovik:1996gx, Rosado:1999yn, Novales-Sanchez:2013rav}. Physically, it determines the correlation between the spin and charge distributions of neutrinos has the same dimensions as that of the charge radius. The anapole term from the general vertex for electromagnetic interactions, $\bar \nu \Lambda_\mu \nu A^\mu$, is defined by \cite{Barroso:1984re, Abak:1987nh,Rosado:1999yn}
\begin{equation}
\Lambda _{\mu }(q)=-\gamma _{\mu }\gamma _{5}F(q^{2}) \simeq -\gamma 
_{\mu}\gamma _{5}q^{2}\textit {\textbf{a}}_{\nu}\,,
\end{equation}
where the form factor ‘$F(q^{2})$’ is related to the neutrino anapole moment ‘$\textit {\textbf{a}}_{\nu \alpha}$’ by the expression,
\begin{equation}
\textit {\textbf{a}}_{\nu} =-\left. \frac{dF_{\nu }(q^{2})}{dq^{2}} \right|_{q^{2}=0} \,.
\end{equation}
By comparing with eq. (\ref{eq:ncrs}), the SM prediction for the neutrino anapole moment can be written in terms of the neutrino charge radius as in the following \cite{Barroso:1984re, Abak:1987nh,Rosado:1999yn, Novales-Sanchez:2013rav, Cadeddu:2018dux},
\begin{equation}
\textit {\textbf{a}}_{\nu_{\rm SM}}=-\frac{\langle r_{\nu }^{2}\rangle _{\rm SM}}{6},
\end{equation} 
and numerical value for the electronic flavor accordingly is given by
\begin{eqnarray}
\textit {\textbf{a}}_{\overline{\nu}_{e \rm SM}}=4.98 \times 10^{-32} \ \ \rm cm^{2}. \nonumber
\end{eqnarray}
For the CE$\overline{\nu}_e$NS with reactor anti-neutrinos, the anapole moment contribution is added to the SM cross-section by replacing the $\sin^2{\theta_W}$ with the effective weak mixing angle in eq. (\ref{eq:gv&ga}) in the following way, 
\begin{eqnarray}
\sin ^{2} \theta _{W}\rightarrow \sin ^{2}\theta _{W} \left(1 - \frac{\pi \alpha_{em}}{18 \sqrt{2} \sin ^{2}\theta _{W} G_F} \textit {\textbf{a}}_{\nu_\alpha} \right).
\label{eq:NCR}
\end{eqnarray}
Again, unlike neutrino millicharges, the anapole moment does not have a direct dependence on the inverse nuclear recoil energy and on the target mass which makes it relatively less sensitive at lower recoil. Now, we fit the parameter $\textit {\textbf{a}}_{\overline{\nu}_{e \rm}}$ with the observed data for the three QFs. The results obtained from this fitting analysis are shown in Fig. \ref{fig:NAM} while the parameter bounds at 90\% C.L. are given in the following

\begin{empheq}[box=\widefbox]{align}
-2.10\times 10^{-30}<\textit {\textbf{a}}_{\overline{\nu}_e}/\rm cm^2<5.10\times 10^{-30}\, \ \ \ \ \ \ \ \ \ \ (\text{Lindhard Model})\  \nonumber \\
-1.90\times 10^{-30}<\textit {\textbf{a}}_{\overline{\nu}_e}/\rm cm^2<4.90\times 10^{-30}\, \ \ (\text{Bonhomme \textit{et al} data})\  \nonumber \\
-1.38\times 10^{-30}<\textit {\textbf{a}}_{\overline{\nu}_e}/\rm cm^2<4.80\times 10^{-30}\, \ \ \ \ \ \ (\text{Sarkis \textit{et al} Model})\  \nonumber \\
-1.15\times 10^{-30}<\textit {\textbf{a}}_{\overline{\nu}_e}/\rm cm^2<4.13\times 10^{-30}\,  \ \ \ \ \ \ \ \ \ \ \ \ \ \ \ \ \ \ (\text{YBe data})\  \nonumber \\
-0.28\times 10^{-30}<\textit {\textbf{a}}_{\overline{\nu}_e}/\rm cm^2<3.26\times 10^{-30}\, \ \ \ \ \ \ \ \ \ \ \ \ \ \ \ \ (\text{Jones data})\  \nonumber \\
-0.14\times 10^{-30}<\textit {\textbf{a}}_{\overline{\nu}_e}/\rm cm^2<3.15\times 10^{-30}\,  \ \ \ \ \ \ \ \ \ \ \ \ \ \ \ \ \ \ \ (\text{Fef data})\
\end{empheq}

Again, Linhard and Sarkis \textit{et al} models give comparable bounds to those obtained before from the coherent scattering using COHERENT data \cite{Khan:2022not} while from the Fef data they are a factor of five stronger than the previous bounds \cite{Khan:2022not}.

\section{\label{sec:concl}Summary and Conclusions}
We have analyzed data from the recently reported observation of the coherent elastic antineutrino- nucleus scattering process with Dresden-II boiling water nuclear reactor neutrinos using the germanium detector. The maximum nuclear recoils produced with reactor antineutrinos can be about 2 keV or below. At such low energy recoils, there is unprecedentedly large uncertainty in the quenching factor. This leads to a large fluctuation in the theoretical prediction of the event energy spectrum in the standard model. Any new physics searches depend on the accuracy of the quenching factor. 
In this regard, we used several theoretical models and data on the quenching factor from various experiments. In the main part of this work, we have shown how different choices of the quenching factor alter the standard model predictions or any new physics sensitivities.

 It was shown that the precise knowledge of the quenching factor is of central importance for the observation of the true signal CE$\overline{\nu}_e$NS in the SM and for any new physics sensitivity. We have included a complete list of the analytical functions of the quenching factor and used them for estimating the value of the weak mixing angle and the neutrino electromagnetic interactions. It is important that these analytical functions could be used for any other new physics models with CE$\overline{\nu}_e$NS from the reactor antineutrinos. It should be noted that there also exist other experimental data for the quenching factor of germanium element, but for nuclear recoils due to the reactor energies, the quenching factors used here are the only relevant ones. We have shown that the Linhard, Sirkis $ et \ al $ model and Bonhomme $et \ al$ are in strong tension with the Jones, iron-filter and photo-neutron data in the regime of nuclear recoils below about 2 keV and a more realistic model is needed. It was consistently shown that the iron-filtered data over-estimate the theoretical predictions of the event energy spectrum while the Linhard model underestimates the predictions which leads to the large disagreement for the value of weak mixing angle and on the bounds of the neutrino electromagnetic parameters.

Using the two theoretical models, namely the Linhard model ~\cite{lindhard1963integral} and the Sarkis $et \ al$ model \cite{Sarkis:2020soy}, and four available data sets for the quenching factor, an older one by Jones $et \ al$ \cite{Jones:1975zze} and the three recent ones with iron-filtered \cite{Colaresi:2022obx} and photo-neutrons and by Bonhomme $et \ al$ 
 \cite{Bonhomme:2022lcz}, we have derived the weak mixing angle and constraints on the possible electromagnetic interactions of neutrinos using the observed data. We have shown how the best fits and the uncertainties on the weak mixing angle and bounds on neutrino millicharge, neutrino magnetic moment, neutrino charge radius and neutrino anapole moment depend on the choice of the quenching factor. The iron-filtered data gives stronger constraints on all parameters comparable to the two theoretical models. Our results for the weak mixing angle and neutrino magnetic moment are with iron-filtered data and photo-neutron data are in agreement with those obtained in ref. \cite{Coloma:2022avw, Liao:2022hno, AristizabalSierra:2022axl} while the constraints on the other electromagnetic interactions obtained here are the first ones.
 
In general, our study provides the first direct limits on electromagnetic properties of neutrinos in addition to the weak mixing angle using the keV scale nuclear recoils due to the reactor antineutrinos. The constraints on the neutrino millicharges are one to two orders of magnitude stronger than the previous bounds from the coherent scattering process using the COHERENT data, while other bounds are comparable to the previous ones. As discussed before these variations are due to the lack of precise knowledge of the quenching factor for pure elements like germanium at low nuclear recoils. We note that neutrino millicharges are more sensitive to the choice of the quenching factor because of the interference of its amplitude with the standard model interactions and its special dependence on the inverse power of the nuclear recoil energy and on the target nuclear mass. We conclude that the precise understanding of the quenching factor is essential for the new physics searches with such low-energy nuclear recoils due to the CE$\overline{\nu}_e$NS.  

\begin{acknowledgments}
\noindent 
I am thankful to professor Juan Collar for providing me with all the necessary data and information about the experiment. I am also thankful to Dr. Y. Sarkis and Dr. B. Cervantes for useful discussions. This work is financially supported by the Alexander von Humboldt foundation.
\end{acknowledgments}

\bibliographystyle{apsrev4-1}
\bibliography{refs}

\end{document}